\documentclass[sigconf,balance=false]{acmart}

\usepackage[T1]{fontenc}
\usepackage{algorithm}
\usepackage{algpseudocode}
\usepackage{amsmath}
\usepackage{amsfonts}
\usepackage{float}

\usepackage{textcomp}
\usepackage{svg}
\usepackage{amsmath}
\usepackage{xcolor}
\usepackage{graphicx}
\usepackage{graphics}
\usepackage{float}
\usepackage{cclicenses}
\usepackage{xspace}
\usepackage{subfigure}
\usepackage{makecell}
\usepackage{multirow} 
\usepackage{wrapfig}
\usepackage{enumitem}

\usepackage{graphicx}      
\usepackage{subfigure}    

\newcommand{\name}{\texttt{FastBioDL}}



\begin{document}
\title{Adaptive Parallel Downloader for Large Genomic Datasets}
%
%
\author{Rasman Mubtasim Swargo}
\affiliation{%
  \institution{Missouri University Science and Technology}
  \country{Missouri, USA}}
\email{rs75c@mst.edu}

\author{Engin Arslan}
\affiliation{%
  \institution{Meta}
  \country{California, USA}}
\email{enginarslan@meta.com}

\author{Md Arifuzzaman}
\affiliation{%
  \institution{Missouri University Science and Technology}
  \country{Missouri, USA}}
\email{marifuzzaman@mst.edu}

%
\begin{abstract}
Modern next-generation sequencing (NGS) projects routinely generate terabytes of data, which researchers commonly download from public repositories such as SRA or ENA. Existing download tools often employ static concurrency settings, leading to inefficient bandwidth utilization and prolonged download times due to their inability to adapt to dynamic network conditions. We introduce \name, a parallel file downloader designed for large biological datasets, featuring an adaptive concurrency controller. \name~frames the download process as an online optimization problem, utilizing a utility function and gradient descent to adjust the number of concurrent socket streams in real-time dynamically. This approach maximizes download throughput while minimizing resource overhead. Comprehensive evaluations on public genomic datasets demonstrate that \name~achieves up to $4x$ speedup over state-of-the-art tools. Moreover, in high-speed network experiments, its adaptive design was up to $2.1x$ faster than existing tools. By intelligently optimizing standard HTTP or FTP downloads on the client side, \name~provides a robust and efficient solution for large-scale genomic data acquisition, democratizing high-performance data retrieval for researchers without requiring specialized commercial software or protocols.

\end{abstract}

\maketitle

\section{Introduction}
The rapid growth of data generated by Next-Generation Sequencing (NGS) technologies has significantly influenced biological and biomedical research. Fundamental public repositories, such as the National Center for Biotechnology Information's (NCBI)~\cite{ncbi} Sequence Read Archive (SRA) and the European Nucleotide Archive (ENA)~\cite{ena}, currently store exabytes of raw sequencing data. These archives form the backbone of modern genomics by facilitating global collaboration and scientific validation. However, the effectiveness of these resources is significantly limited by the efficiency of data retrieval processes.

The massive scale of contemporary sequencing projects introduces substantial practical challenges for data acquisition. Tools and protocols designed for smaller datasets are architecturally insufficient for transferring the terabyte- and petabyte-scale data now commonly encountered. Researchers regularly face extended download times, unpredictable transfer failures, and inefficient use of costly institutional network infrastructure. This "last-mile" data delivery challenge not only delays research but also poses barriers for researchers at institutions with fewer resources, complicating efforts to maintain computational reproducibility.

An example of this inadequacy is found in the tools provided by major data repositories. The widely used fastq-dump~\cite{sra-toolkit} utility from the SRA Toolkit, originally single-threaded, proves inadequate for large modern datasets, severely underutilizing available processing cores and network bandwidth. Although the introduction of tools like fasterq-dump~\cite{sra-toolkit} improved multithreading capabilities, their concurrency levels remain static, configured only once at the start of transfers. This approach fails to address the dynamic nature of modern network environments. While libraries like pysradb~\cite{pysradb} partially address these constraints, downloading large datasets remains a lengthy process. Thus, more robust and efficient solutions are clearly necessary.

While specialized UDP-based accelerators like IBM Aspera FASP~\cite{aspera} can saturate wide-area links, their need for licensed server software and institutional deployment effort makes them inaccessible to most researchers. However, our work demonstrates that a real-time concurrency optimizer can leverage plain socket connections to fully utilize the large bandwidth available in modern high-performance networks. This open-source approach democratizes high-performance data transfer for all researchers, ultimately accelerating scientific discoveries that rely on large-scale genomic downloads.


In this paper, we introduce \name, a new parallel downloader specifically developed to efficiently and reliably retrieve large biological datasets from public repositories using standard HTTP or FTP protocols. The major contributions of our work include:

\begin{itemize}
    \item \textbf{High-Performance Downloader:} A downloader explicitly designed for large biological datasets, utilizing parallel HTTP streams to effectively download data from repositories such as ENA and NCBI. Achieving up to 4x speedup compared to the widely used prefetch tool (SRA Toolkit).
    \item \textbf{Adaptive Concurrency Control:} A dynamic mechanism that adjusts the number of concurrent download operations in real-time based on current network and system performance, optimizing overall throughput and resource utilization. In high-speed network scenarios, it can provide up to 2.1x transfer speed compared with traditional fixed concurrency methods.
\end{itemize}

\section{Related Work}
The Sequence Read Archive (SRA)~\cite{SRA} and the European Nucleotide Archive (ENA)~\cite{ena} have become important resources for life science research because of the increase in Next-Generation Sequencing (NGS) data. These repositories store raw sequencing data mainly in a compressed format called SRA. Researchers access and analyze these data for their projects. Because the data files are very large, several downloading tools have been developed to increase efficiency. Each tool has its own strengths and limitations.

The National Center for Biotechnology Information's (NCBI)~\cite{ncbi} SRA Toolkit~\cite{sratools} is one of the most widely used software tools. Previously, fastq-dump was the standard tool for downloading and converting .sra files into the widely used FASTQ format. However, fastq-dump is single-threaded, making it slow and inefficient for large sequencing runs. To address this issue, fasterq-dump was introduced. Fasterq-dump uses multiple threads and handles temporary files more efficiently, significantly reducing conversion time. The SRA Toolkit also includes a tool called prefetch, which reliably downloads sra files and supports resuming interrupted downloads. Using prefetch before conversion is often recommended. Without using prefetch first, fasterq-dump does not download data efficiently. Although users can control the concurrency through parameters, the target users typically do not have expertise in managing complex network settings or handling dynamic network conditions. This remains a limitation of fasterq-dump.

pysradb~\cite{pysradb} provides an interface for searching metadata and downloading data using underlying tools such as the SRA Toolkit. Although it supports parallel downloads, the concurrency level is static; it must be set before execution starts. Thus, the concurrency limitation seen with fasterq-dump remains. Tools like parallel-fastq-dump~\cite{pfastq-dump} focus on speeding up the conversion of compressed .sra files to FASTQ format through parallelization. However, they usually require the entire .sra file to be fully downloaded first before parallel conversion can begin.\\
IBM Aspera's~\cite{aspera} FASP protocol is an example of a specialized high-speed file-transfer method. It employs advanced techniques to maximize bandwidth and reduce the impact of network conditions. Typically, it provides the fastest file transfer speeds for very large files, provided both the source and destination have the Aspera client/server software installed and properly configured. However, because Aspera uses non-standard protocols and requires specific software setups on both ends, it is less universally accessible than standard HTTP-based methods. \name~targets the far larger community that must rely on vanilla HTTP endpoints maintained by ENA and NCBI, and shows that intelligent client‑side optimization can still unlock multi‑gigabit speeds.\\  
Several studies have addressed this issue from the perspective of general file transfers. Transport-layer approaches have received considerable attention, especially in wide-area data transfers. Examples of such methods include congestion control algorithms like QTCP~\cite{qtcp}, BBR~\cite{BBR}, and PCC-Vivace~\cite{vivace}. Among these, BBR demonstrates better performance compared to TCP Cubic, even under random packet-loss conditions. However, since these approaches operate at the network level, they often fail to overcome performance bottlenecks caused by limitations in I/O operations.\\
The primary challenge with application-layer methods is the large search space and the slow evaluation process involved. Notable methods include concurrent file transfers~\cite{concurrency}, pipelining techniques~\cite{pipelining}, employing multiple Data Transfer Nodes (DTNs)~\cite{dtn}, and using parallel TCP streams~\cite{parallel}.\\
Probing techniques are effective for addressing various network-related issues, such as throughput prediction~\cite{osti_10324560,10824943} and bandwidth estimation~\cite{probing}. However, probing can introduce extra network traffic, potentially leading to congestion.\\
Globus~\cite{globusGlobus} is a popular data transfer service that estimates optimal concurrency settings. However, Globus typically underestimates concurrency to prevent over-utilization, which can lead to slower transfer speeds. Recent studies~\cite{prasanna,concurrency} have modeled configuration selection as an online convex optimization problem. Falcon uses standard optimization techniques like gradient descent to optimize the concurrency. Other tools, like FDT~\cite{fast_data_transfer}, mdtmFTP~\cite{The_MDTM_project}, and Marlin~\cite{marlin}, handle network and I/O tasks separately.

In this study, we introduce \name, a parallel downloader designed specifically to address the challenges of downloading large biological datasets. It uses multiple concurrent HTTP streams. Importantly, \name~features adaptive concurrency, dynamically adjusting the number of active threads based on real-time network and system conditions. This approach optimizes transfer speeds and provides a flexible alternative to traditional static concurrency settings and specialized transfer protocols.

\section{Motivation}
\begin{figure}[ht]
    \centering
    \includegraphics[width=0.4\textwidth]{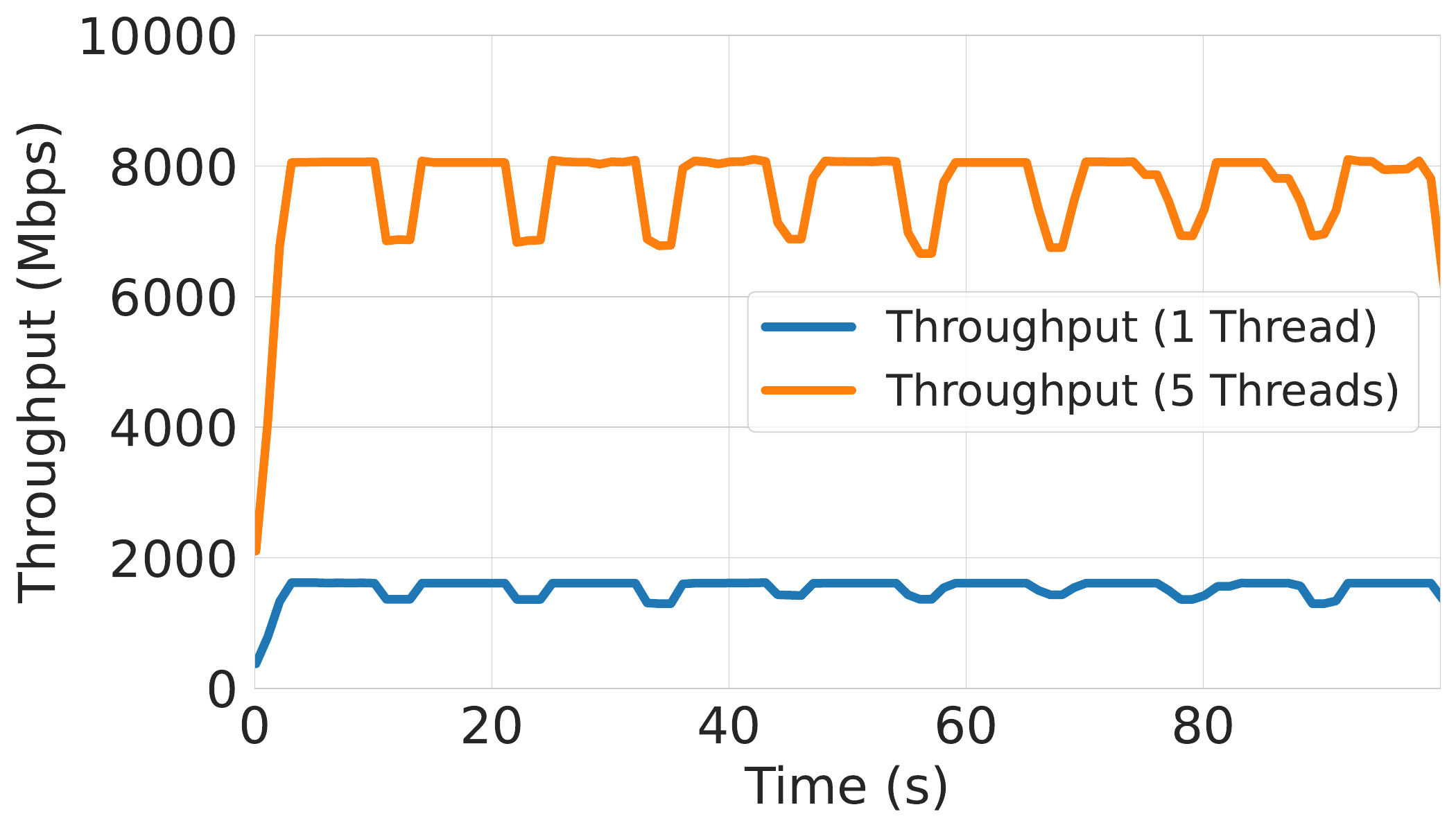}
    \caption{Single-threaded FTP downloads underutilize network bandwidth, as measured by the iperf3 tool.}
    \Description{A graph showing that single-threaded FTP downloads do not fully utilize the available network bandwidth, as measured by iperf3.}
    \label{fig:singleThread}
    \vspace{-3mm}
\end{figure}

The widely-used SRA/FASTQ downloader, fastq-dump, downloads data from the NCBI repository using a single-threaded HTTPS connection. This approach was sufficient until datasets reached petabyte scales. Sequentially downloading and converting files became impractical and introduced significant bottlenecks. Figure \ref{fig:singleThread} illustrates how a single-threaded download fails to utilize the available network bandwidth efficiently.

\begin{figure}[ht]
    \centering
    \includegraphics[width=0.4\textwidth]{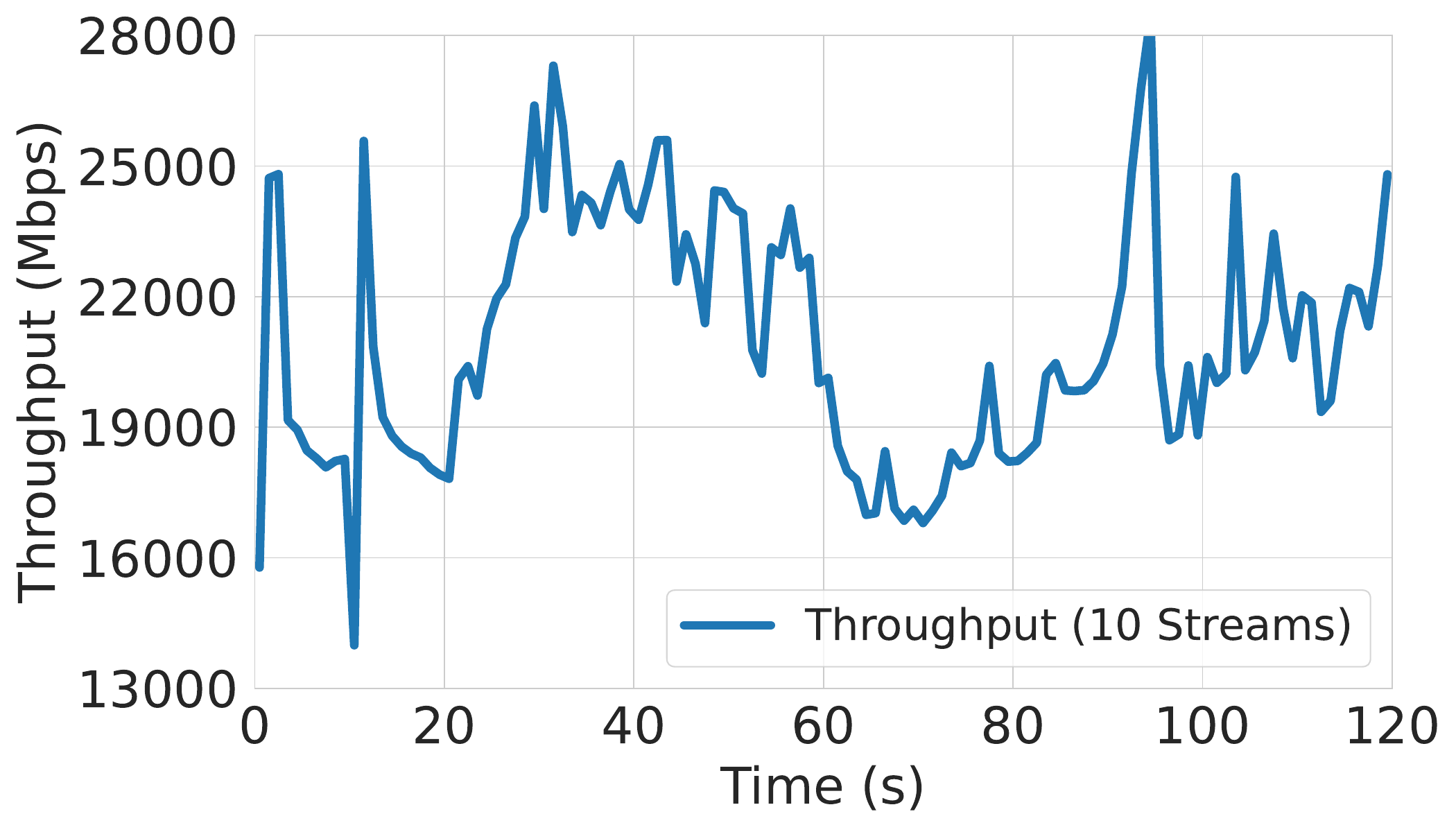}
    \caption{Real-world network throughput is inherently dynamic, as demonstrated by iperf3 measurements over a two-minute interval. Static concurrency settings may lead to suboptimal resource utilization.}
    \Description{Real-world network throughput is inherently dynamic, as demonstrated by iperf3 measurements over a two-minute interval. Static concurrency settings may lead to suboptimal bandwidth utilization.}
    \label{fig:iperf3}
    \vspace{-3mm}
\end{figure}

To improve concurrency and resource utilization, NCBI introduced a newer tool called fasterq-dump. Typically used alongside the prefetch tool, fasterq-dump introduces concurrency during the conversion stage from the downloaded file to FASTQ format. However, prefetch itself still downloads files sequentially. To achieve parallel downloads of multiple SRA files, users often rely on external scripts or job management systems. Even then, concurrency must be fixed at the beginning of the process.  In reality, the available bandwidth between a researcher's machine and a public data archive is a volatile resource, subject to fluctuations from network congestion, competing background traffic, and server-side load. Any fixed concurrency level is therefore guaranteed to be suboptimal for the majority of a transfer's duration. Static concurrency can cause either underutilization of bandwidth or excessive load. As shown in Figure \ref{fig:iperf3}, network throughput can vary significantly even within short periods, making static concurrency settings inefficient.\\
To address these challenges systematically, we model data downloading as an online optimization problem. The objective is to find the optimal concurrency strategy, \(C(t)\), that maximizes the average throughput, \(T\), over the duration of the transfer, \(D\). This can be expressed as
\[
\max_{C(t)} \frac{1}{D} \int_0^{D} T(C(t),t)\, dt
\]
Here, the instantaneous throughput \(T\) is explicitly shown as a function of both the chosen concurrency level \(C(t)\) and other time‑dependent network conditions. In a static system, \(C(t)\) is a constant. In an adaptive system, \(C(t)\) is the control variable adjusted by a continuous feedback loop. The goal of the \name~adaptive engine is to continuously solve for the optimal \(C(t)\) that maximizes the objective function. This formal, optimization‑centric approach provides the theoretical foundation for a more robust, efficient, and resilient data downloader.

\section{\name: Systems Design}
\begin{figure*}[ht]
\centering
\includegraphics[width=0.99\textwidth]{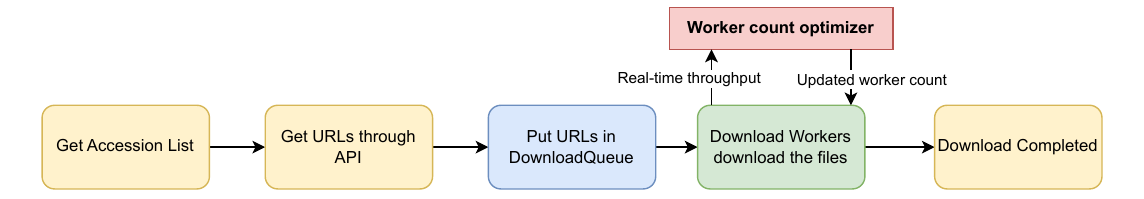}
\caption{\name~downloads the SRA accessions with adaptive concurrency.}
\Description{\name~downloads the SRA accessions with adaptive concurrency.}
\label{fig:workflow}
\vspace{-3mm}
\end{figure*}

\name~is designed to integrate concurrency optimization into the biological data downloading process. An accession number represents a unique alphanumeric identifier assigned to individual data records within biological sequence and molecular data repositories. The downloader first reads an accession list comprising multiple runs, which can be retrieved from repositories such as NCBI or ENA. It then generates download links utilizing APIs provided by either the ENA Portal API or the NCBI E-utilities API. Once URLs are obtained, they are queued for download, and parallel download workers are initiated. Concurrently, dedicated threads monitor and report real-time throughput data to the optimizer. The optimizer aggregates throughput information over a specified probing duration and dynamically adjusts concurrency levels to balance performance and overhead. Figure \ref{fig:workflow} illustrates the complete workflow. In the following sections, we discuss the implementation details.

\begin{table}[ht]
    \centering
    \begin{tabular}{|c|c|c|}
    \hline
    K & Avg Download Speed (Mbps) & Avg Concurrency \\
    \hline
    1.01 & 701.2 & 6.77 \\
    1.02 & 815.8 & 6.23 \\
    1.05 & 743.9 & 4.64 \\
    \hline
    \end{tabular}
    \vspace{2mm}
    \caption{Penalty coefficient K balances concurrency overhead and achieving convergence.}
    \label{tab:setup}
    \vspace{-5mm}
\end{table}

\subsection{Utility Function}
To optimize concurrency levels effectively, the optimizer requires a suitable utility function. This function is inspired by several recent online transfer optimization studies~\cite{drl,concurrency}. We adapted it specifically for file downloading tasks rather than generalized bi-directional file transfers. Our utility function prioritizes maximum throughput while using the minimum number of threads, thereby minimizing overhead caused by additional threads. The utility function is defined as:

\begin{align*}
U(throughput, concurrency) &= \frac{throughput}{k^{concurrency}}
\end{align*}

This utility function rewards increased throughput while penalizing higher concurrency through the penalty constant $k$. The function ensures concurrency levels rise only when throughput improvements sufficiently justify the associated concurrency overhead, guiding the optimizer toward an optimal balance of high download speeds and efficient resource utilization. Because we want to maximize this utility, but our implementation uses optimizers like gradient descent (which finds minima), we minimize the negative utility in code.

Selecting an appropriate value for $k$ requires a comprehensive evaluation. Higher values of $k$ increase the penalty term, discouraging rapid increases in concurrency. Conversely, lower values of $k$ allow faster convergence by promoting more aggressive concurrency increases. The challenge arises when performance deteriorates due to excessive concurrency overhead. To visualize the impact of K on the utility function, we show the mathematical implications. Assuming an infinite bandwidth network, concurrency level $C$ and a fixed per-thread throughput $\alpha$, we can define the utility function as,
\[
U(C)=\frac{\alpha\,C}{k^{C}}, 
\qquad C\in\{1,\dots,C_{\max}\},\;k>1,
\]
Setting $\partial U/\partial C=0$ yields
\[
C^{\star}= \tfrac{1}{\ln k},\qquad 
U''(C^{\star})<0,
\]

Here, $C^{\star}$ denotes the unique global maximum for concurrency within the interval $[1, C_{\max}]$, where the negated utility function has a unique global minimum, making it unimodal. 
Although the negated utility function is not a convex function globally, gradient-based methods initialized reasonably can reliably converge to the optimal concurrency in this simplified model.
Theoretically, if the optimal bandwidth utilization requires a concurrency level exceeding $C^{\star}$, our approach may fall short. As $k$ determines the converged concurrency level's upper limit, we need to select this parameter carefully. However, given the capabilities of current and emerging HPC networking infrastructures that support bandwidths of several hundred gigabits per second, this limitation is not yet a practical concern. It also shows that in practice, k is one of the key factors that determines the optimal concurrency that~\name will try to achieve.
Table \ref{tab:setup} illustrates the impact of k during data downloads. $k=1.02$ yields the highest download speed; $k=1.01$ suffers from overhead due to overly aggressive concurrency, whereas $k=1.05$ 
fails to fully utilize available bandwidth due to overly conservative concurrency adjustments. Based on these findings, we selected $k=1.02$ for subsequent experiments. However, k is a tunable parameter, and users or system admins can override the default value to set the level of aggressiveness of the optimizer.

\subsection{Adaptive Concurrency Optimization}
\name~employs an optimizer to dynamically adjust the concurrency level during download operations. A probing function periodically collects network metrics such as throughput for a fixed duration (default is set at 3 seconds) using the concurrency value determined by the optimizer. During these probing durations, instantaneous throughput data are logged and aggregated. The utility function computes utility values based on aggregated throughput data, providing feedback to the optimizer. The optimizer subsequently updates concurrency settings at the end of each probing duration using current utility values and historical parameter adjustments. This continuous feedback loop dynamically adjusts concurrency in real-time to ensure optimal download performance.

\begin{algorithm}[H]
\caption{\name~Optimizer Thread}
\label{alg:optimizer_concise}
\begin{algorithmic}[1]
\Require Shared Throughput Logs, Shared Process Status Arrays, Configuration (Optimization Method, Probing Time)
\Ensure Dynamic updates to Shared Process Status Arrays

\State Initialize Optimizer state and initial concurrency levels.

\Comment{Select candidate concurrency levels}
\While{Transfer not fully complete}
    \State OptimalConcurrency $\gets$ SelectBest(Candidates, Scores)
    \State Set Worker Statuses to OptimalConcurrency.
    \State Run for Probing Time.
    \State Measure Throughput from Logs.
    \State Evaluate Performance Score.
\EndWhile

\Comment{Ensure workers stop on exit}
\State Set all Worker Statuses to 0.
\end{algorithmic}
\end{algorithm}

\begin{figure*}[htbp]
    \centering
    \subfigure[Gradient Descent]{%
        \includegraphics[width=0.45\textwidth]{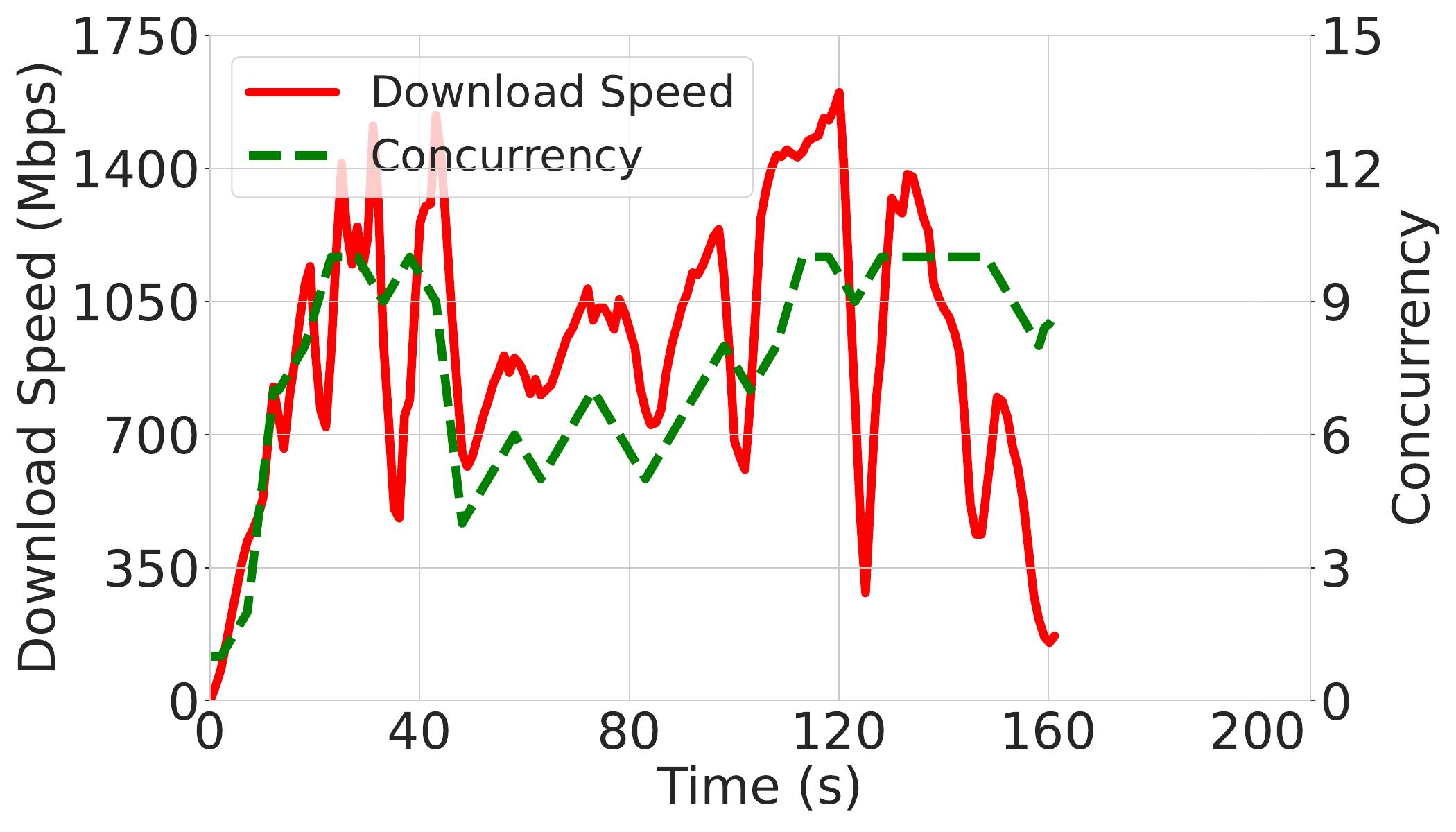}%
        \label{fig:eval_opt_gd}%
    }
    \quad
    \subfigure[Bayesian Optimizer]{%
        \includegraphics[width=0.45\textwidth]{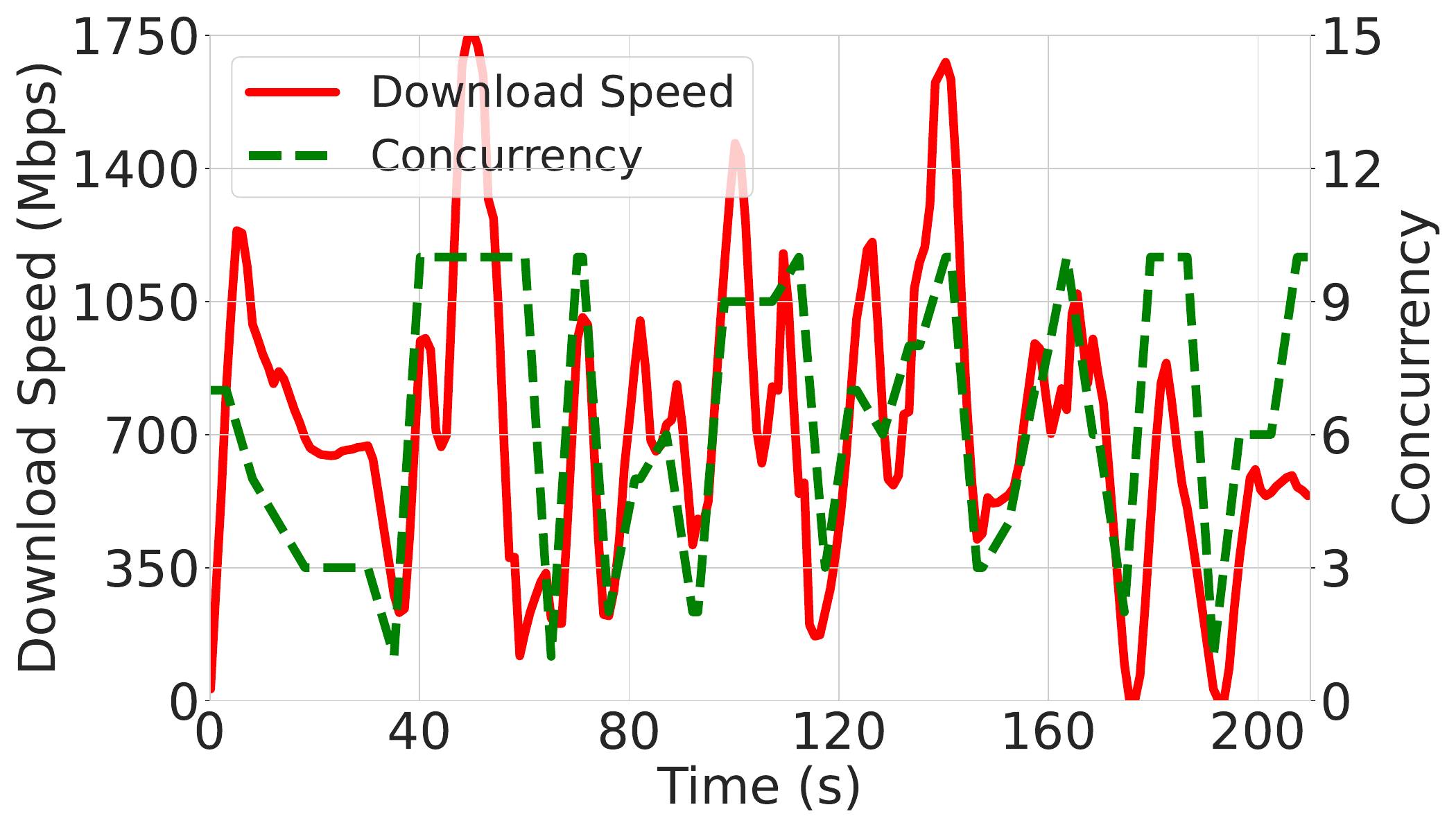}%
        \label{fig:eval_opt_bayes}%
    }
    \vspace{-2mm}
    \caption{Gradient Descent outperforms Bayesian Optimizer as it suffers from constantly evolving systems dynamics.}
    \Description{}
    \label{fig:eval_opt}
\end{figure*}

We evaluated two widely used optimization techniques: gradient descent and Bayesian optimization. Gradient descent incrementally adjusts concurrency based on throughput feedback, making minor iterative changes. Bayesian optimization, on the other hand, selects concurrency levels through probabilistic sampling. Gradient descent demonstrated superior performance compared to Bayesian optimization because its incremental adjustments minimized interruptions during ongoing downloads. Gradient descent makes small, local moves and relies only on the very recent throughput readings, so it does not use a model at all. On the other hand, Bayesian optimization must first seed a Gaussian surrogate model with a few random trials; when early samples arrive during momentary disk or network spikes as demonstrated in Figure~\ref{fig:iperf3}, the model is skewed. That bad fit leads the acquisition function to pick thread counts that are far from the optimal setting for that moment, forcing big jumps and socket resets. Each jump feeds more noisy data back into the model, so the correction cycle drags on. In practice, the surrogate never stabilized within a single run, and total copy time stayed about 20\% slower than gradient descent, as illustrated in Figure~\ref{fig:eval_opt}, which is an average of five runs. Because the cost of a wrong guess is high and the signal is noisy, we find that gradient descent is the better choice for the downloader.

\section{Evaluation}
We first conducted a competitive analysis against existing open-source genomics data downloading tools. For this experiment, we used production HTTPS API endpoints from NCBI and ENA. Subsequently, we leveraged NSF FABRIC testbed~\cite{fabric-2019} to demonstrate the effectiveness of \name~in next-generation high-speed networks.


\begin{table*}[htbp]
\centering
\begin{tabular}{@{}l l l r c c@{}}
\toprule
\textbf{Alias (this article)} & \textbf{BioProject ID} & \textbf{Organism / Sample type} & \textbf{Files taken} & \textbf{Total\ size} & \textbf{Size range} \\
\midrule
Breast‑RNA‑seq   & PRJNA762469 & \textit{Homo sapiens} (breast transcriptome) & 10 & 22.06 GB & 1.72 – 3.03 GB \\
HiFi‑WGS        & PRJNA540705 & \textit{Homo sapiens} (PacBio long‑read WGS)  & 6  & 56.15 GB & 8.10 – 10.81 GB \\
Amplicon‑Digester & PRJNA400087 & Anaerobic digester metagenome                & 43 & 1.91 GB  & 13.43 – 66.47 MB \\
\bottomrule
\end{tabular}
\caption{Summary of evaluation datasets used in this study}
\label{tab:datasets}
\vspace{-5mm}
\end{table*}

\begin{table}[htbp]
\centering
\begin{tabular}{@{}llcc@{}}
\toprule
\textbf{Dataset} & \textbf{Tool} & \textbf{Concurrency} & \textbf{Speed (Mbps)} \\
\midrule
\multirow{3}{*}{Breast‑RNA‑seq} 
 & prefetch & $3.00 \pm 0.00$ & $517.70 \pm 40.12$ \\
 & pysradb  & $8.00 \pm 0.00$ & $749.32 \pm 141.82$ \\
 & \name~& $3.42 \pm 0.62$ & $989.12 \pm 92.35$ \\
\midrule
\multirow{3}{*}{HiFi‑WGS} 
 & prefetch & $3.00 \pm 0.00$ & $246.82 \pm 18.97$ \\
 & pysradb  & $8.00 \pm 0.00$ & $220.56 \pm 82.67$ \\
 & \name~& $4.92 \pm 0.21$ & $594.75 \pm 50.52$ \\
\midrule
\multirow{3}{*}{Amplicon-Digester} 
 & prefetch & $3.00 \pm 0.00$ & $29.15 \pm 3.53$ \\
 & pysradb  & $8.00 \pm 0.00$ & $29.10 \pm 2.17$ \\
 & \name~& $4.14 \pm 0.42$ & $117.47 \pm 2.03$ \\
\bottomrule
\end{tabular}
\caption{Mean concurrency and download speed (Mbps) with standard deviation for the three evaluation datasets}
\label{tab:perf}
\vspace{-5mm}
\end{table}

\subsection{Comparison with State-of-the-Art}

We compared \name~with widely used existing solutions, employing a probing duration of 5 seconds, and Gradient Descent as the optimizer. Specifically, \name~was evaluated against pysradb and the SRA Toolkit's prefetch tool. These experiments were executed on Google Colab with 12 GB RAM.
To evaluate \name~under a spectrum of workload characteristics, we chose three public BioProjects, each mapped to a distinct alias (Table \ref{tab:datasets}).
\textbf{Breast‑RNA‑seq} (PRJNA762469) comprises ten human breast Illumina RNA‑seq runs with per‑file sizes in the 1.72–3.03 GB range (average $\approx$ 2.4 GB). These mid‑sized short‑read files test the optimizer's ability to adapt when both connection setup time and sustained throughput matter.
\textbf{HiFi‑WGS} (PRJNA540705) contains six PacBio HiFi whole‑genome reads, each 8.10–10.81 GB (average $\approx$ 9.5 GB); the large, continuous objects stress bandwidth utilization and reveal any ceiling effects of high concurrency.
\textbf{Amplicon‑Digester} (PRJNA400087) provides 43 amplicon libraries from an anaerobic digester metagenome, with very small files (13.4–66.5 MB, average $\approx$ 40 MB). Here the workload is dominated by connection churn and metadata overhead.
Using small, medium, and large datasets drawn from different sequencing technologies ensures that performance claims generalize across file‑size patterns and I/O profiles.\\
The fastq-dump tool downloads larger, uncompressed FASTQ files and consequently exhibits considerably slower performance compared to the other three methods, highlighting inefficiencies stemming from single-threaded downloads and on-the-fly conversions for large datasets. Consequently, it was not compared to the other tools. The SRA Toolkit's prefetch (part of the fasterq-dump pipeline) tool employs some internal parallelism, downloading files using a static concurrency level of three threads. pysradb allows users to specify thread counts; we chose eight threads for our experiments, as it is the most commonly used. Conversely, \name~utilizes adaptive concurrency. The latter three tools download compressed SRA Lite files.
Accurately evaluating performance is challenging due to variability in network conditions. To address this issue, each experiment was conducted five times using a round-robin approach. Table~\ref{tab:perf} summarizes the results; values are reported as mean $\pm$ standard deviation.\\



\begin{figure}[ht]
    \centering
    \includegraphics[width=0.45\textwidth]{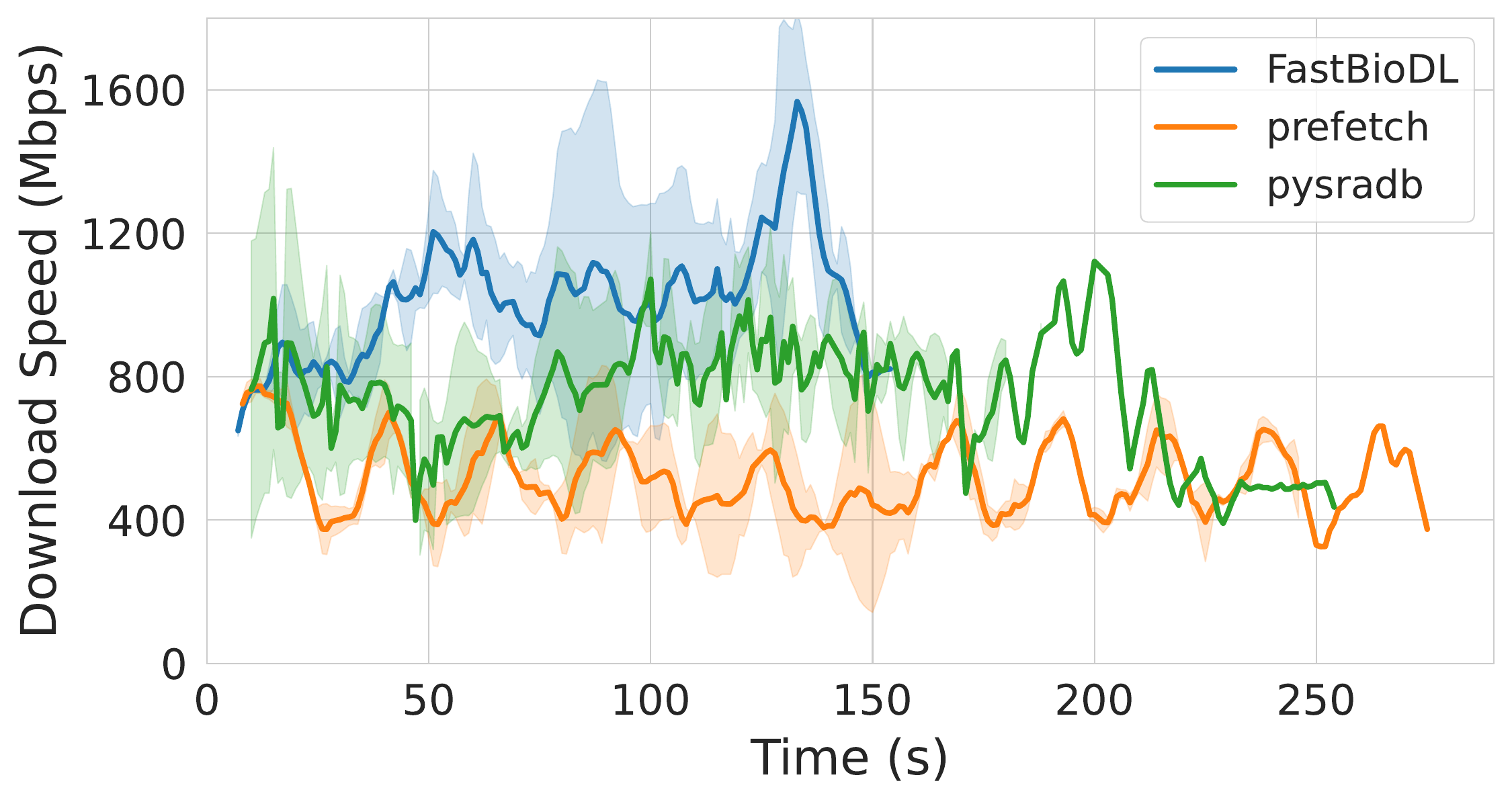}
    \vspace{-2mm}
    \caption{\name~demonstrates superior performance, demonstrating higher throughput and faster completion times compared to existing tools.}
    \Description{\name~demonstrates superior performance, demonstrating higher throughput and faster completion times compared to existing tools.}
    \label{fig:eval_compare}
    \vspace{-5mm}
\end{figure}

\begin{figure*}[htbp]
  \centering

  \begin{minipage}[b]{0.32\textwidth}
    \centering
    \textbf{Bandwidth: 10000 Mbps\\Speed per Thread: 500 Mbps\\File Size: 100GB}
  \end{minipage}\hfill
  \begin{minipage}[b]{0.32\textwidth}
    \centering
    \textbf{Bandwidth: 10000 Mbps\\Speed per Thread: 1400 Mbps\\File Size: 100GB}
  \end{minipage}\hfill
  \begin{minipage}[b]{0.32\textwidth}
    \centering
    \textbf{Bandwidth: 20000 Mbps\\Speed per Thread: 1400 Mbps\\File Size: 512GB}
  \end{minipage}

  \begin{minipage}[b]{0.32\textwidth}
    \centering
    \includegraphics[width=\linewidth]{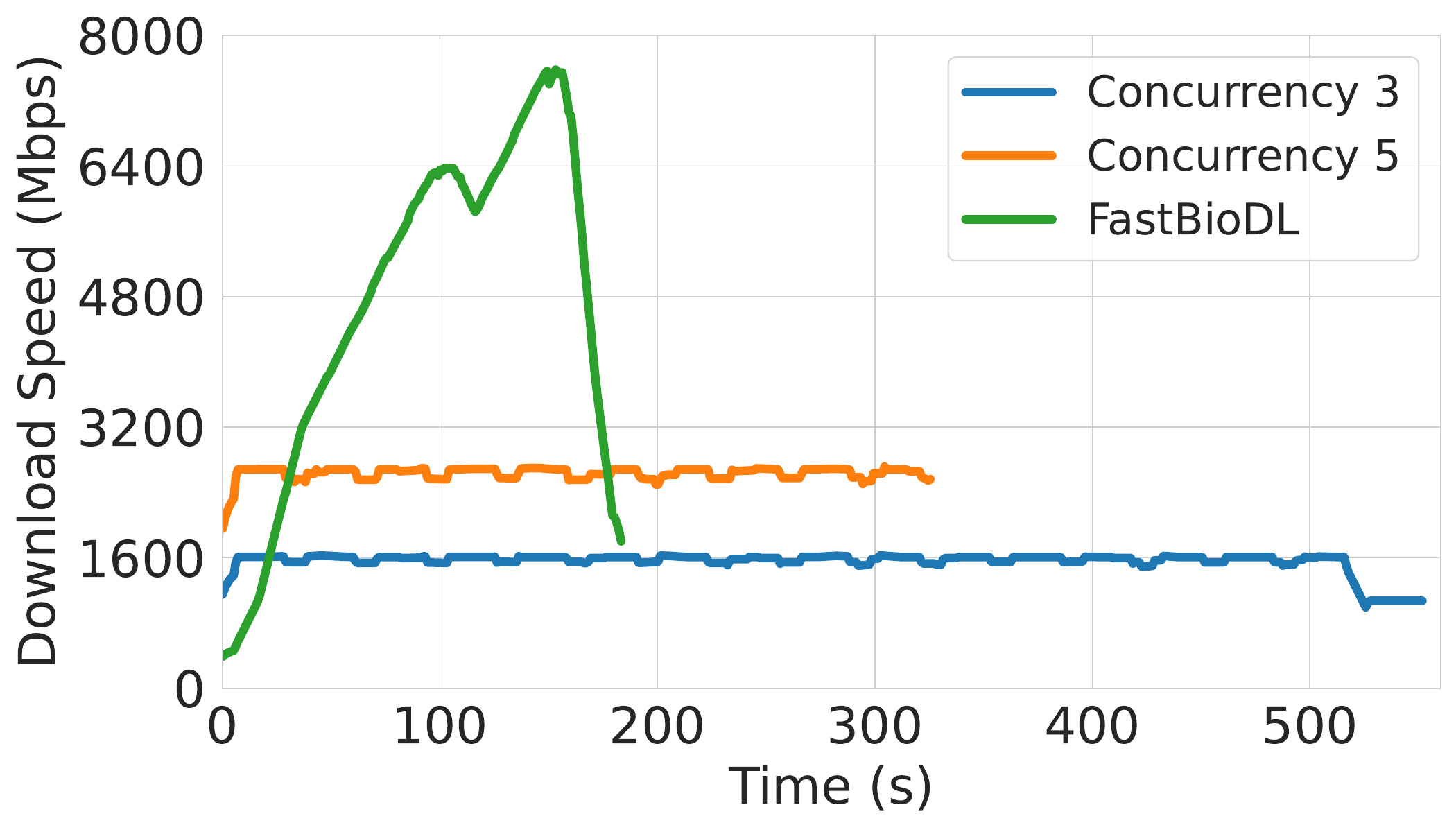}
  \end{minipage}\hfill
  \begin{minipage}[b]{0.32\textwidth}
    \centering
    \includegraphics[width=\linewidth]{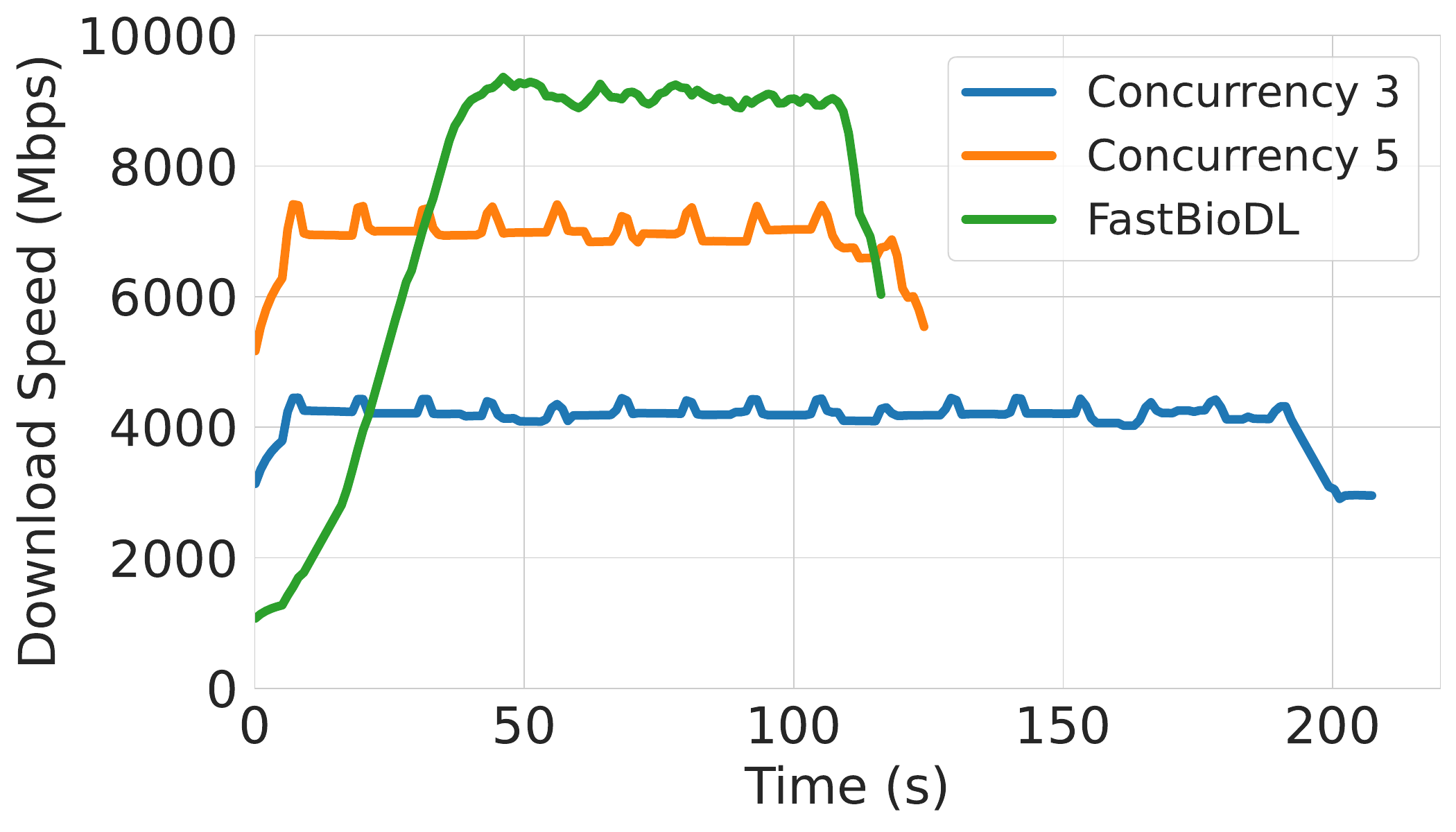}
  \end{minipage}\hfill
  \begin{minipage}[b]{0.32\textwidth}
    \centering
    \includegraphics[width=\linewidth]{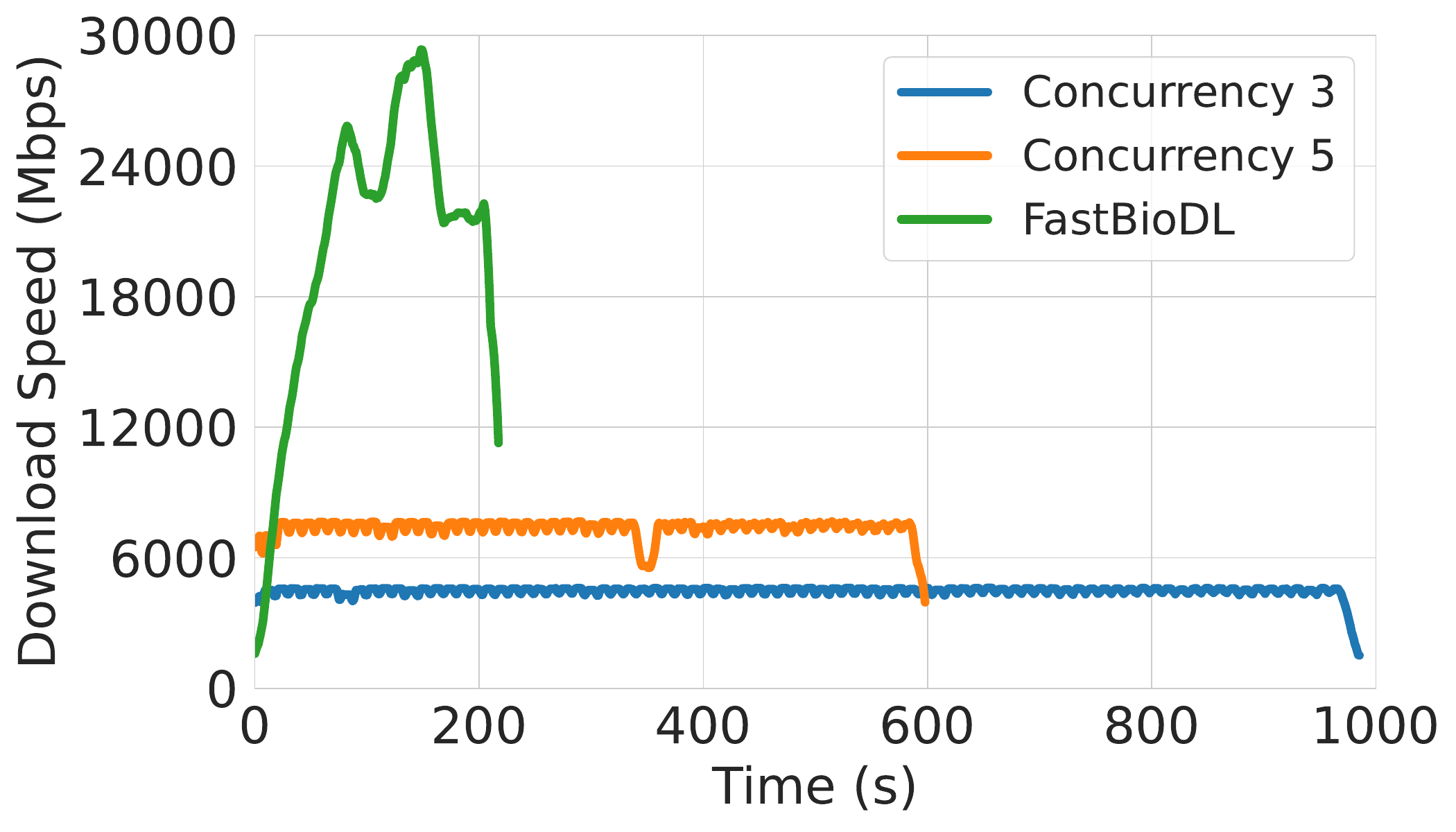}
    \vspace{-6mm}
  \end{minipage}

  \caption{Comparison of download speeds using \name's adaptive concurrency versus fixed concurrency levels in high-speed network scenarios. \name's optimized concurrency rapidly achieves optimal performance, resulting in improved throughput and better resource utilization.}
  \Description{Comparison of download speeds using \name's adaptive concurrency versus fixed concurrency levels in high-speed network scenarios. \name's optimized concurrency rapidly achieves optimal performance, resulting in improved throughput and better resource utilization.}
  \label{fig:comparison}
\end{figure*}

In all of the datasets, \name~outperformed the existing tools. In the Breast-RNA-seq dataset, our solution tops prefetch ($\approx1.9\times$) and pysradb ($\approx1.3\times$) in download speed. Although pysradb was close, it employed more than double the concurrency of \name, thus introduces much more resource overhead. In HiFi-WGS dataset \name~chooses more threads because this is necessary to achieve better speedup: approximately $2.4\times$ and $2.7\times$ compared to prefetch and pysradb, respectively. Our solution performed the best in the Amplicon-Digester dataset too. It got $\approx4x$ speedup compared to both tools. 


Figure~\ref{fig:eval_compare} plots the per‑second mean throughput and its 68\% confidence band for \name, prefetch, and pysradb on the Breast‑RNA‑seq dataset. It shows the instantaneous behavior of the downloader tools.
In these particular trials, \name~achieved approximately 1800 Mbps peak throughput, while other tools reached up to 1400 Mbps. \name~completed downloads in 160 seconds, which is 38\% and 43\% faster than pysradb and prefetch, respectively. These results demonstrate that \name’s adaptive parallel downloading strategy provides a more efficient solution for retrieving large biological datasets compared to the existing tools.

\subsection{Performance on Next-Generation Networks}
Another key advantage of \name~is its purposeful design for next-generation networks. Existing tools rely on static concurrency values, making them incapable of scaling to the higher bandwidths available in modern HPC environments. To demonstrate this limitation, we set up an FTP server, stored several hundred gigabytes of randomly generated files, and conducted downloads from another host over a high-speed link. Our experiments were conducted on the Fabric testbed~\cite{fabric-2019}, spanning two HPC centers, NCSA and SALT. The hosts used in these experiments were configured with 64 GB RAM, Dell Express Flash P4510 1TB SFF storage, and NVIDIA Mellanox ConnectX-6 NIC. All tests utilized NVMe drives as both the source and destination storage.

To evaluate \name's efficiency and adaptability across varying systems and network configurations, we designed three experimental scenarios by throttling both network bandwidth and per-thread download speeds. For the first two experiments, we used 100 GB random files; for the third, 512 GB files. We compared \name~against fixed concurrency levels of 3 and 5, the most common values used by existing tools. Note that these tools could not be used directly in our experiment, as they only support downloads from preexisting genomic repositories.

Figure~\ref{fig:comparison} shows the mean instantaneous speeds for each approach. In the first scenario, with total bandwidth limited to 10,000 Mbps and per-thread speed capped at 500 Mbps, the theoretical optimal concurrency was 20. \name~logs showed an average of around 10 threads. This is because the optimizer starts with one thread and probes every 5 seconds; by the time it reached the optimal concurrency, the download had already completed, thus lowering the mean concurrency. Nevertheless, \name~completed the download 44\% faster than the fixed concurrency level of 5 and 67\% faster than level 3, achieving around 7,500 Mbps throughput, whereas the fixed levels experienced significantly more underutilization.

In the second scenario, we increased per-thread speed to 1,400 Mbps, reducing the theoretical optimal concurrency to approximately 7. \name~averaged around 6 concurrent threads. Although concurrency level 5 finished only 8 seconds behind \name, it failed to fully utilize the available bandwidth. \name~achieved a throughput of roughly 9,300 Mbps, compared to 7,300 Mbps for concurrency 5. In the third scenario, we allowed full bandwidth utilization of the testbeds ($\approx$20 Gbps), with per-thread speed still at 1,400 Mbps. Here, \name~approached full use of the available bandwidth, averaging 14 concurrent threads (the theoretical optimum being 14.3). Even under this high-performance setting, \name~outperformed fixed concurrency strategies, offering speedups of $1.3x$and $2.1x$ over fixed concurrency levels of 5 and 3, respectively. These results clearly demonstrate that \name~adapts effectively to high-bandwidth environments and will significantly outperform static-concurrency tools as faster networking becomes more prevalent in both HPC systems and consumer applications. 


\section{Conclusion}
We introduce \name, a parallel data mover designed to address the critical bottleneck of large-scale data acquisition in genomics, thereby accelerating scientific discoveries. With rigorous evaluation we demonstrate that an adaptive and online optimization approach for downloading large dataset yields substantial performance improvements over the current widely used static methods. In production scenarios, \name~reduced download completion times by up to 43\% and increased average throughput by up to $4\times$ compared to the standard NCBI SRA Toolkit. In high-speed testbed environments designed to highlight the impact of network variability and high bandwidth availability, \name's adaptive design was up to $2.1\times$ faster than a comparable fixed-concurrency downloader. These results show that adapting to real-time systems and networking conditions is a fundamentally more efficient and robust strategy for large-scale data transfer. By enhancing data retrieval speed and reliability, \name~lowers the barriers to conducting large-scale computational experiments. Future work includes integrating \name~seamlessly with Genomic data processing workflows, extending adaptive control design to additional transfer protocols beyond HTTP and FTP, and adding more features such as on-demand and pipelining downloads.


\bibliographystyle{plain}
\bibliography{references0}

\end{document}